\documentclass[prb,twocolumn,amsmath,amssymb,aps,longbibliography,superscriptaddress,citeautoscript,notitlepage, bibnotes,nofootinbib]{revtex4-2}

\pdfoutput=1

\usepackage{braket}
\usepackage{bm}

\usepackage{graphicx}

\usepackage[normalem]{ulem}
\usepackage{subfigure}
\usepackage{textcomp}
\usepackage{graphicx}
\usepackage{amssymb}
\usepackage{amsmath}
\usepackage{bm}
\usepackage{amsmath, amsthm, amssymb}
\usepackage{dsfont}
\usepackage[colorlinks=true,linkcolor=blue,urlcolor=blue,citecolor=blue]{hyperref}

\DeclareMathOperator{\Tr}{\mbox{Tr}}

\usepackage{pdfpages} 
\makeatletter
\AtBeginDocument{\let\LS@rot\@undefined}
\makeatother

\newcommand{\nag}{{\phantom{\dagger}}}

\def\ket#1{\mathinner{|{#1}\rangle}}
\def\braket#1{\mathinner{\langle{#1}\rangle}}

\def\beq{\begin{equation}}
\def\eeq{\end{equation}}
\def\bea{\begin{eqnarray}}
\def\eea{\end{eqnarray}}

\begin{document}
\renewcommand{\vec}[1]{{\boldsymbol{#1}}}

	\title{Theory of competing excitonic orders in insulating  WTe$_2$ monolayers}
		\author{Yves H. Kwan}
	\affiliation{Rudolf Peierls Centre for Theoretical Physics,  Clarendon Laboratory, Oxford OX1 3PU, UK}
	\author{T. Devakul }
	\affiliation{Department of Physics, Princeton University, Princeton, New Jersey 08540, USA}
	\author{S. L. Sondhi}
	\affiliation{Department of Physics, Princeton University, Princeton, New Jersey 08540, USA}
		\author{S. A. Parameswaran}
	\affiliation{Rudolf Peierls Centre for Theoretical Physics,  Clarendon Laboratory, Oxford OX1 3PU, UK}

	\date{\today}
	
\begin{abstract} 
We develop a theory of the excitonic phase recently proposed as the zero-field insulating state observed near charge neutrality in monolayer WTe$_2$. Using a Hartree-Fock approximation, we numerically identify two distinct gapped excitonic phases: a spin density wave state for weak non-zero interaction strength and spin spiral order at stronger interactions, separated by a narrow window of non-excitonic quantum spin Hall insulator. We  introduce a simplified model capturing key features of the WTe$_2$ band structure, in which these phases appear as distinct valley ferromagnetic orders. We link the competition between the excitonic phases to the orbital structure of  electronic wavefunctions at the Fermi surface and hence its  proximity to the underlying gapped Dirac point in WTe$_2$. We briefly discuss collective modes of the two excitonic states, and comment on implications for experiments.
 \end{abstract}
	
\maketitle
	
	\section{Introduction}
	
	When the ground state  of a semimetal or narrow-gap semiconductor  
	becomes unstable to electron-hole Coulomb attraction, it is replaced by  an equilibrium condensate of electron-hole pairs (excitons)~\cite{halperin1968short,halperin1968long,jerome1967,kohn1967,cloizeaux1965,zittartz1967,kunes2015}. This new excitonic state of matter is typically insulating, but separated by a phase transition from a conventional band insulator. Although theoretically proposed over half a century ago,  the excitonic state has proven to be remarkably elusive experimentally, with significant progress towards this goal only coming in the past decade or so~\cite{Eisenstein,kogar2017,du2017,cercellier2007,wakisaka2009,seki2014,lu2017,fukutani2019}.	
		
	Recent transport and tunneling measurements on ultraclean monolayers of the transition-metal dichalcogenide WTe$_2$ have been argued to be consistent with an excitonic insulating ground state near the charge neutrality point~\cite{WTe2exciton,WTe2Landau}. Signatures of this state develop only at low temperatures, indicating an electron ordering transition. Strikingly, despite being  insulating at zero field~\cite{tang2017,fei2017,wu2018,WTe2exciton,WTe2Landau}, WTe$_2$  shows robust Shubnikov-de Haas  oscillations at high magnetic fields~\cite{WTe2Landau}. This suggests that the insulator is either highly unconventional, or else transitions to a conductor with increasing field. Improving the understanding of the zero-field insulating state is a necessary first step to exploring this intriguing system. 

		 An obstacle to this goal is posed by the complex band structure of WTe$_2$ which, in the absence of interactions,  consists of a pair of over-tilted Dirac cones~\cite{muechler2016}, weakly gapped by spin-orbit coupling (SOC). This leads to a pair of conduction band minima (electron pockets) at an incommensurate wavevector $\pm \vec{q_c}$, flanking a single valence band maximum (hole pocket) at the Brillouin zone $\vec{\Gamma}$ point. The anisotropic pockets, strong SOC, the twofold `valley' index labelling electron pockets, and possibly nontrivial orbital structure on the Fermi surface (FS) due to the near-Dirac band structure are in stark contrast to the simplified starting point that, with few exceptions~\cite{wu2015,ataei2020},  underlies  studies of the excitonic state. A  theory of excitonic insulators in WTe$_2$ must incorporate these complexities and clarify their role in influencing its phase structure.
		 
		 	\begin{figure}[!t]
	\includegraphics[width=1\linewidth]{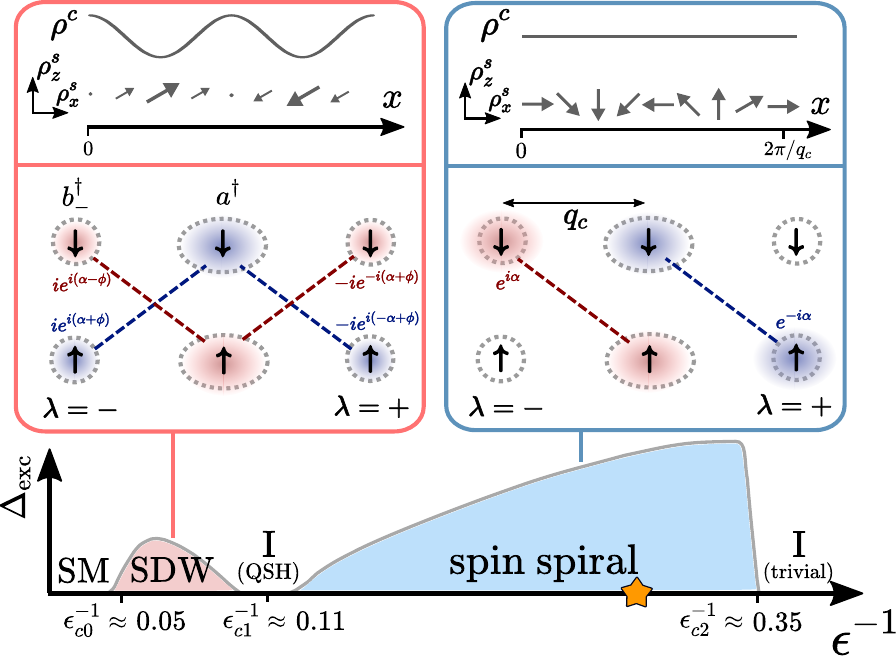}
	\caption{Bottom: Hartree-Fock phase diagram and evolution of excitonic pairing scale $\Delta_{\text{exc}}$  with  interaction strength $\epsilon^{-1}$ (star indicates estimated experimental value). Excitonic order is present (absent) in the  SDW and spin spiral (SS)  (semimetal (SM) and insulator (I)) phases. Top: charge/spin order and schematic pairing structure for SDW/SS. Fermi pockets of the non-interacting $k\cdot p$ model are schematically shown with dotted grey lines.}
	\label{fig:PhaseDiagram}
\end{figure}

  Here, we explore the  phase diagram of WTe$_2$, focusing on  spin and valley pseudospin order in the excitonic states and its interplay with the orbital structure of the energy bands.  We first map out the phase diagram numerically (Fig.~\ref{fig:PhaseDiagram}) within a Hartree-Fock (HF) treatment of interactions.  We find two {distinct} excitonic insulators, corresponding to spin-density wave (SDW) and spin spiral (SS) orders, for different interaction strengths. We introduce a simplified  analytically tractable model that captures the  low-energy structure of the interaction-renormalized  bands in WTe$_2$. This gives an  intuitive picture where individual SDW/SS excitons are degenerate, but compete due to exciton interactions. We link this competition to the orbital content of the conduction band, which can be tuned experimentally.
    We sketch qualitative  features of the SDW/SS collective modes and discuss their experimental signatures. We close by speculating on   possible implications for   
    high-field transport.

\section{$\textrm{WTe}_2$ Model}
We begin with a  $\bm{k}\cdot\bm{p}$ theory~\cite{WTe2exciton,qian2014} valid near the  WTe$_2$ $\bm{\Gamma}$-point, $H_0 = \sum_{\bm{k}}h_{\alpha\beta}(\bm{k}) c^\dagger_{\bm{k}\alpha}c^\nag_{\bm{k}\beta}$, where $\alpha, \beta$ are composite spin-orbital indices, and
\bea\label{eq:H0}
\!\!\hat{h}(\bm{k}) \!=\!\varepsilon_+(\bm{k})\!+\![\varepsilon_-(\bm{k})\!+\!{\delta}]\tau^z  +v_xk_x\tau_xs_y+v_yk_y\tau_ys_0.\,\,\,\,
\eea
 The Pauli matrices $\tau^\mu$, $s^\mu$  act in  orbital  and spin space with $\tau^z=\pm1$  ($s^z = \pm1$) referring to $d$, $p$ orbitals ($\uparrow, \downarrow$ spins) respectively, $\varepsilon_{\pm}(\bm{k}) =  \frac{1}{2}\left(\varepsilon_d(\bm{k}) \pm \varepsilon_p(\bm{k})\right)$ where $\varepsilon_d(\bm{k}) =  a\bm{k}^2+b\bm{k}^4$ and $\varepsilon_p(\bm{k}) =-\frac{\bm{k}^2}{2m}$, with  $a=-3~\text{eV\r{A}}^2$, $b=18~\text{eV\r{A}}^4$, $m=0.03~\text{eV$^{-1}$\r{A}}^{-2}$, $v_x=0.5~\text{eV\r{A}}$, $v_y=3~\text{eV\r{A}}$ are chosen to match the {\it ab initio} band structure of Ref.~\cite{WTe2exciton}. $\delta$ controls the band overlap, with $\delta<0$ ($\delta>0$) corresponding to a semimetal (semiconductor). We take $\delta=-0.45~\text{eV}$ which sets the Fermi energy ($E_F$) of the noninteracting bands at charge neutrality to $E_F\simeq -0.493~\text{eV}$, yielding a hole pocket at  $\bm{\Gamma}$, and two electron pockets with minima at $\bm{q_c}=\pm 0.3144\,$\r{A}$^{-1}\bm{\hat{x}}$, incommensurate with the reciprocal lattice vector $G_x=1.81\,$\r{A}$^{-1}$. 
 
$H_0$ respects parity $\hat{P}=\tau_z$ and time-reversal $\hat{T}=is_y\hat{K}$ symmetries ($\hat{K}$ is complex conjugation), and hence its bands are twofold degenerate. Absent SOC ($v_x=0$), $H_0$ has \emph{overtilted} Dirac cones at $\bm{q}_D=\pm 0.2469\,$\r{A}$^{-1}\hat{\bm{x}}$. SOC  gaps the Dirac point, and yields an indirect negative band gap; however it retains $U_s(1)$ spin rotation symmetry about the $y$ axis, that we assume henceforth.

We now rewrite interactions, which are density-density in spin and orbital space, in the band eigenbasis defined by diagonalizing~\eqref{eq:H0}, $H_0 = \sum_{\bm{k}n\sigma} \varepsilon_{\bm{k}}^n d^\dagger_{\bm{k} n \sigma} d^\nag_{\bm{k} n \sigma}$. Here, $d^\dagger_{\bm{k} n \sigma}  = \sum_{\alpha}  u_{\bm{k}n\sigma}^{\alpha} c^\dagger_{\bm{k}\alpha}$ where the sum is over spins/orbitals and $u_{\bm{k}n\sigma}^{\alpha}$ are the relevant Bloch functions,  which we make diagonal in spin space by choosing the  spin quantization axis along $y$, and $n=a,b$ correspond to valence and conduction bands. In this basis, we find
\bea\label{eq:Hint}
H_{\text{int}}&=&\frac{1}{2 A}\sum_{\bm{q}}U(\bm{q}) :\!\rho^\dagger_{\bm{q}}\rho^\nag_{\bm{q}}\!:,
\eea
where  $A$ is the system area, $:\!\ldots\!:$ denotes normal ordering with respect to the Fock vacuum, and $U(\bm{q}) =\frac{e^2}{2\epsilon\epsilon_0 q}\tanh\frac{q\xi}{2}$ is a dual-gate screened interaction.  [Experiments correspond to a screening length $\xi = 25\,$nm, and relative permittivity $\epsilon\simeq 3.5$.] We consider $\epsilon^{-1}$ as a parameter that tunes the overall strength of interactions. 
Eq.~\eqref{eq:Hint} introduces the band-projected densities $\rho_{\bm{q}} \equiv \sum_{nn'\bm{k}\sigma} F^{nn';\sigma}_{\bm{k}-\bm{q},\bm{k}} d^\dagger_{n\sigma, \bm{k} -\bm{q}} d^\nag_{n'\sigma \bm{k}}$
and form factors $F^{nn';\sigma\sigma'}_{\bm{k},\bm{k}'}=\braket{u_{\bm{k}n\sigma} | u_{\bm{k}'n'\sigma'}}$. Apart from constraints imposed by $\hat{P}$ and $\hat{T}$ (which respectively require $F^{nn';\sigma}_{\bm{k}\bm{k}'}=F^{nn';\sigma}_{-\bm{k},-\bm{k}'}$ and $F^{nn';\sigma}_{\bm{k}\bm{k}'}=[F^{nn';\bar{\sigma}}_{-\bm{k},-\bm{k}'}]^*$), the latter can generically  vary as the bands traverse the BZ. 

\section{Hartree-Fock phase diagram}
The Hamiltonian $H = H_0 + H_{\text{int}}$ defined by \eqref{eq:H0} and \eqref{eq:Hint} captures the key features of WTe$_2$ relevant to studying its low-energy behaviour near charge neutrality. We numerically study the phase diagram via self-consistent HF calculations with momentum cutoffs $|k_x|<\frac{3q_c}{2},|k_y|<\frac{G_y}{4}$, where $G_y=1.01$\,\r{A}$^{-1}$ is the reciprocal lattice vector in the $y$ direction. Anticipating a possible  excitonic instability, we allow for translational symmetry breaking at wavevector $\bm{q_c}$. More details of the self-consistent Hartree Fock calculations can be found in Appendix~\ref{app:HF}.

Fig.~\ref{fig:PhaseDiagram} shows the phase diagram as a function of the interaction strength $\epsilon^{-1}$. In the non-interacting limit, the system starts off with three Fermi pockets, as sketched by the grey dotted lines in the middle row of Fig.~\ref{fig:PhaseDiagram}. As $\epsilon^{-1}$ is increased, the system remains semimetallic until $\epsilon^{-1}_{c0}\approx0.05$ where it transitions into a gapped SDW phase. This phase possesses non-trivial excitonic ordering, diagnosed by the integrated $\bm{q}=\bm{q_c}$ coherence $\Delta_\text{exc}\equiv\sqrt{\sum_{\alpha,\beta} |\langle c^\dagger_{\bm{k}\alpha}c^{\phantom{\dagger}}_{\bm{k}+\bm{q_c}\beta} \rangle|^2}$ and exhibits both SDW order in the $xz$ spin plane (orthogonal to the SOC axis) at wavevector $\bm{q_c}$, and charge density (CDW) wave order at $2\bm{q_c}$ (see Fig.~\ref{fig:orderparams}). The SDW preserves combined $\hat{P}\hat{T}$ symmetry, so its bands remain doubly degenerate. 
Excitonic order is suppressed in a small window around $\epsilon^{-1}_{c1}\approx0.11$ in favor of a non-excitonic quantum spin Hall insulator, which yields to a second excitonic $\hat{P}\hat{T}$-broken phase that we dub the spin spiral (SS). Unlike the SDW, SS has no CDW order (at least at purely electronic level), and the local spin polarization is of constant magnitude and rotates in the $xz$ spin-plane with wavevector $\bm{q_c}$. Finally, for $\epsilon^{-1} \geq \epsilon^{-1}_{c2}\approx 0.35$, $\Delta_\text{exc}$ vanishes and the system is a non-excitonic trivial insulator. For the experimental interaction strength $\epsilon^{-1}\approx 0.28$, the ground state is a SS exciton insulator with an indirect gap of $\sim 230$~meV (see Fig.~\ref{fig:U0bandstructures}) and local spin polarization of $\sim 0.002\,\mu_{\rm{B}}\text{\r{A}}^{-2}$. [Enforcing $\hat{P}$ and $\hat{T}$ gives a single excitonic  phase with slightly higher energy \cite{WTe2exciton, bab-comm}.]

At large $\epsilon^{-1}$, $\Delta_\text{exc}$ receives contributions from $\bm{k}$-states across much of the folded BZ, suggesting that  it distorts the bands even far from 
$E_F$. However throughout the SDW and in the SS near $\epsilon^{-1}_{c1}$,  excitonic coherence is localized in $\bm{k}$-space around the centers of the Fermi pockets. This suggests that  we can understand excitonic coherence and competition between the SDW/SS phases in this regime by focusing only on states near $E_F$.

\section{Effective model}
We now construct a simplified effective model that captures low-energy features of the {renormalized} symmetry-preserving band structure most relevant to the excitonic order. The use of self-consistent HF bands (without excitonic distortion) as a starting point is crucial: since \eqref{eq:Hint} is normal-ordered with respect to the Fock vacuum, bands will naturally deform due to self-exchange when the $\bm{k}\cdot\bm{p}$ model is half-filled (consistent with charge neutrality). Therefore it is only sensible to consider the interplay of the band structure with excitonic condensation  \emph{after} incorporating such renormalization effects. Consequently, the parent state for the excitonic phases is an insulator with a small indirect band gap (even though we began with a semimetal at $\epsilon^{-1}=0$).

\begin{figure}[!t]
	\includegraphics[width=0.9\linewidth,clip=true]{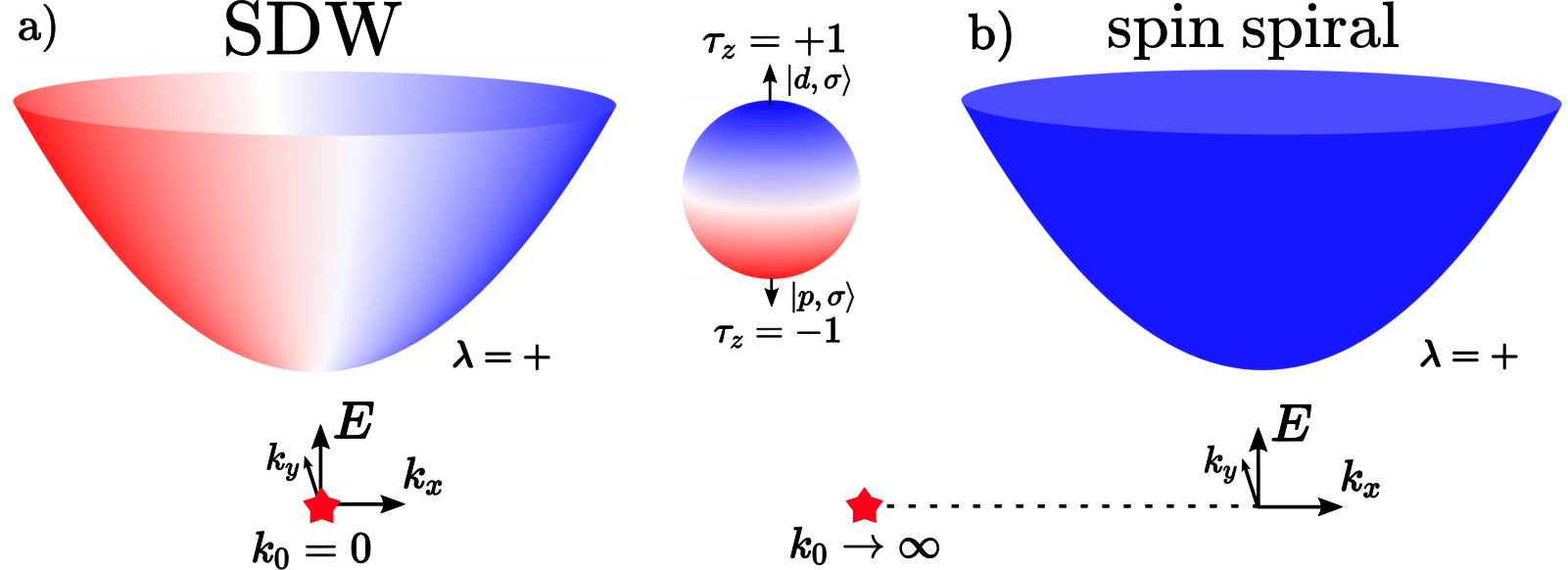}
	\caption{Sketch of $\lambda=+$ valley dispersion and orbital structure for the two limits of~\eqref{eq:valleybloch} appropriate to  (a) SDW and (b) SS phases; red star marks location of gapped Dirac point.}
	\label{fig:BlochStructure}
\end{figure}

Since we are in the regime where excitonic pairing is only peaked around $\bm{\Gamma}$ and $\pm\bm{q_c}$, we restrict attention to Bloch states within three small `pockets' centered about these special momenta~\cite{halperin1968long}. We treat electrons from the two conduction band minima as separate species distinguished by a `valley' pseudospin $\lambda=+,-$, and describe them using independent creation operators $b^\dagger_{\bm{k}\sigma\lambda}$, where $\bm{k}$ is measured from $\lambda\bm{q_c}$. We also introduce $a^\dagger_{\bm{k}\sigma}$, an {electron} creation operator for the valence band maximum. The momentum takes all values within some pocket cutoff $|\bm{k}|<k_\text{cut}$ large enough to encompass the region of excitonic pairing. We approximate the dispersions  by best-fit effective-mass parabolas $\epsilon_{\bm{k}}^{a,b}$ at extrema of the self-consistent bands at $\epsilon^{-1}$ (ignoring small `teardrop' corrections to the band structure). We model the form factors of the valleys as arising from a gapped Dirac cone, with the Dirac point displaced by some wavevector $-\lambda k_0\hat{\bm{x}}$ from the minimum of the dispersion to capture the tilted structure relevant to WTe$_2$. Accordingly, in our model we take the  valley-$\lambda$ Bloch function $\ket{\tilde{u}^\lambda_{\bm{k}b\sigma}}$ at $\bm{k}$ (see Fig.~\ref{fig:BlochStructure}) to be the positive eigenvector of
\begin{equation}\label{eq:valleybloch}
h_{\sigma,\lambda}(\bm{k})=\lambda \tilde{v}_x(k_x+\lambda k_0)\tau_z+\tilde{v}_y k_y\tau_y+\lambda\sigma \tilde{m}\tau_x.
\end{equation}
Though valence band Bloch functions can be modeled similarly in principle, in practice they do not affect the SDW-SS competition. 
The effective  Hamiltonian,
\begin{equation}\label{eq:Heff}
H_{\text{eff}}=\sum_{\bm{q}}\epsilon_{\bm{q}}^aa^\dagger_{\bm{q}\sigma}
a^\nag_{\bm{q}\sigma}\!+
\!\sum_{\lambda}\epsilon_{\bm{q}}^b b^\dagger_{\bm{q}\sigma\lambda} b^\nag_{\bm{q}\sigma\lambda} + \frac{U(\bm{q})}{2A}\!:\!{\rho}^\dagger_{\bm{q}} \rho_{\bm{q}}\!:,
\end{equation}
is normal ordered with respect to the filled valence band.

The natural hierarchy of inter- and intra-pocket interactions $U(q_c)/U(0)\simeq0.025$ admits a physically intuitive separation of scales~\cite{halperin1968long}. In the dominant term approximation (DTA), we retain only the band dispersion and interaction terms 
with small intra-pocket momentum transfers $q\ll q_c$ (form factor effects are negligible {\it within}  each pocket since we can choose a smooth gauge where $F^{nn;\sigma}_{\bm{k},\bm{k}+\bm{q}}\rightarrow 1$ as $q\rightarrow0$). At this order, we determine the existence of an excitonic instability and the momentum structure of excitonic pairing. The DTA has an enhanced $U(4)$ symmetry and hence does not distinguish between exciton phases with distinct spin and valley orders. This degeneracy is resolved at beyond-DTA (bDTA) level, where the neglected $\bm{q}\sim \bm{q}_c,2\bm{q}_c$  interactions split the various states, with the orbital structure of the bands playing a crucial role. The influence of orbital structure on energetic competition is evident already in the few-exciton problem about the insulating state where the valence (conduction) bands of $H_{\text{eff}}$ are filled (empty). As demonstrated in Appendix~\ref{app:twoexcitons}, single excitons with the symmetries of SS/SDW are degenerate, but are split at two-exciton level by inter-valley $2\bm{q}_c$ interactions of their constituent electrons, which are sensitive to  orbital structure via  form factors.

\section{Variational states}
 To fully explore excitonic order, we  consider the  extended HF states (generalizing~\cite{halperin1968long})
\begin{equation}\label{eq:varansatz}
\ket{\Phi} = \prod_{\bm{k},\sigma} \alpha_{\bm{k}\sigma}^\dagger\ket{0},\,\alpha_{\bm{k}\sigma} =  u_{\bm{k}}a_{\bm{k}\sigma} + v_{\bm{k}}\sum_{s\lambda} w^{\sigma}_{s\lambda} b_{\bm{k}s\lambda},
\end{equation}
where the parameters are chosen to minimise $\langle H_\text{eff}\rangle_{\Phi}$ with $u_{\bm{k}}, v_{\bm{k}}$ real and even in $\bm{k}$, $u_{\bm{k}}^2+ v_{\bm{k}}^2=1$,
$\sum_{s\lambda}w^{\sigma}_{s\lambda}\bar{w}^{\sigma'}_{s\lambda}=\delta_{\sigma\sigma'}$, and overbar denotes complex conjugation. Eq.~\eqref{eq:varansatz} describes a state obtained by first folding the BZ by $q_c$, so that all three pockets are centred  at $\bm{\Gamma}$, and then introducing excitonic coherence between the valence band and a specific spin-valley combination  in the conduction band parametrized by $w$. At  DTA level, the energy is  $w$-independent, and $u_{\bm{k}},v_{\bm{k}}$ (which set the momentum structure of excitonic coherence) are determined by self-consistently solving the coupled integral equations 
$\sqrt{2} v_{\bm{k}}=\left(1-{\xi_{\bm{k}}}/{\sqrt{\xi_{\bm{k}}^2+\Delta_{\bm{k}}^2}}\right)^{1/2}$,  
$\Delta_{\bm{k}} =\sum_{\bm{k}'}U(\bm{k}-\bm{k}')g_{\bm{k}'}$, 
where $g_{\bm{k}} = u_{\bm{k}}v_{\bm{k}}$, $\xi_{\bm{k}}=\frac{1}{2}(\bar{\epsilon}^b_{\bm{k}}-\bar{\epsilon}^a_{\bm{k}})$, with  $\bar{\epsilon}^{a}_{\bm{k}} = {\epsilon}^{a}_{\bm{k}} +  \sum_{\bm{k}'}U(\bm{k}-\bm{k}')v_{\bm{k}'}^2$ and $\bar{\epsilon}^{b}_{\bm{k}} = {\epsilon}^{b}_{\bm{k}} - \sum_{\bm{k}'}U(\bm{k}-\bm{k}')v_{\bm{k}'}^2$. For small $\epsilon^{-1}$ in the exciton phases of Fig.~\ref{fig:PhaseDiagram}, we find that $g_{\bm{k}}~\sim\langle a b^\dagger \rangle$, which is a direct measure of excitonic pairing, qualitatively matches the momentum-resolved contributions to $\Delta_{\text{exc}}$ in the HF, justifying the use of the effective model. The $w$-dependence is restored upon perturbatively evaluating the bDTA splitting terms: 
\begin{eqnarray}\label{eq:bDTA}
\delta E[w]&=&D|\Tr W^{+-}|^2-\sum_{ss'}J_{ss'}W^{++}_{ss'}W^{--}_{s's}+\tilde{\mathcal{E}}[w],\\
D&=&U(2\bm{q}_c)|\sum_{\bm{k}}v_{\bm{k}}^2\mathcal{F}^{\uparrow}_{\bm{k}\bm{k}}|^2 = U(2\bm{q}_c)|\sum_{\bm{k}}v_{\bm{k}}^2\mathcal{F}^{\downarrow}_{\bm{k}\bm{k}}|^2,\label{eq:D}\\
J_{ss'}&=&\frac{1}{2}\sum_{\bm{k}\bm{k}'}v_{\bm{k}}^2v_{\bm{k}'}^2U(\bm{k}-\bm{k}'+2\bm{q}_c)[\mathcal{F}^{s^*}_{\bm{k}'\bm{k}}\mathcal{F}^{s'}_{\bm{k}'\bm{k}}+\text{c.c}],\,\,\label{eq:J}
\end{eqnarray}
where $W^{\lambda\lambda'}_{ss'}\equiv\sum_{\sigma} w^{\sigma}_{s\lambda}w^{\sigma^*}_{s'\lambda}$ and $\mathcal{F}^s_{\bm{k}\bm{k}'}\equiv\braket{\tilde{u}^-_{\bm{k}bs}|\tilde{u}^+_{\bm{k}'bs}}$, the sole potentially nontrivial  form factor in our model, depends on the valley Bloch parametrization~\eqref{eq:valleybloch}. Terms in $\tilde{\mathcal{E}}[w]$ give identical energies for SDW/SS; as these are not central to our discussion we relegate them to Appendix~\ref{app:bDTA}.

We now identify the SDW and SS phases in terms of $w$. As shown in Fig.~\ref{fig:PhaseDiagram}, the in-plane SDW is described by
\begin{equation}\label{eq:SDW-w}
\text{SDW:}\,\,\,w^{\sigma}_{s +}=\frac{ie^{i\alpha}}{\sqrt{2}}
\begin{pmatrix}
0&e^{-i\phi}\\e^{i\phi}&0
\end{pmatrix}_{\sigma s},\,\,\,
w^{\sigma}_{s -}=\bar{w}^{s}_{\sigma+},
\end{equation}
which has spin density $\bm{\rho}^s(\bm{r})\sim \sin(q_cx-\alpha)[\sin\phi,0,\cos\phi]$, charge density $\rho^c(\bm{r})\sim \cos[2(q_cx-\alpha)]$ and bDTA splitting $\delta E=D-\frac{J_{\uparrow\uparrow}}{2}$. (Though Ref.~\cite{halperin1968long}  identifies SDW is the ground state for $\mathcal{F}\to 1$, we find that in reality SS is  favored in this limit.) In the SS,
\begin{equation} \label{eq:SS-w}
\text{SS:}\,\,\,w^{\downarrow}_{\uparrow +}=e^{-i\alpha},\,\,\,
w^{\uparrow}_{\downarrow -}=e^{i\alpha}
\end{equation}
with spin density $\bm{\rho}^s(\bm{r})\sim [\sin(q_cx+\alpha),0,\cos(q_cx+\alpha)]$ and 
$\delta E=0$. Inversion exchanges valleys, yielding a spiral with opposite handedness.
[Up to global rotations/translations, the most general $xz$-spin order is an {\it elliptic} spiral,   
$\bm{\rho}^s(\bm{r})\sim [\cos\chi \sin(q_cx),0,\sin\chi\cos(q_cx)]$. As discussed in Appendix~\ref{app:elliptical}, assuming $H$ has $S^y$-rotation symmetry, this generates  a CDW with Fourier component $\rho^c(2\bm{q}_c)\sim |\bm{\rho}^s(\bm{q}_c)|^2\propto \cos(2\chi)$, but 
here energetics force $\chi\to\pm\pi/4, 0$, corresponding to a circular SS with no CDW, or a pure SDW without spiral order~\cite{Zachar} -- the cases we consider.]

Using \eqref{eq:SDW-w} and \eqref{eq:SS-w} in \eqref{eq:bDTA}, we find that the competition between SDW and SS is tuned by $D$ and $J_{\uparrow\uparrow}$: as in our two-exciton warmup problem (Appendix~\ref{app:twoexcitons}), these describe $|\bm{q}|\sim 2q_c$ interactions between the valleys. $D$ is a Hartree term that directly penalizes $2\bm{q}_c$ CDW order, while  $J$ reflects intervalley exchange. SS is unaffected by both contributions due to its perfect spin-valley locking, but the SDW is energetically favored if $2D<J_{\uparrow\uparrow}$.
The central role played by the orbital structure/form factors is apparent in two limiting cases of \eqref{eq:valleybloch}. For $k_0\rightarrow \infty$, where the Dirac physics is invisible to low-lying conduction electrons near the band minimum (recall $E_F$ is in the renormalized band gap)
that participate in excitonic pairing, the Bloch functions are uniform and identical in both valleys, because $h_{\sigma,\lambda}(\bm{k})\sim \tau_z$. Since $\mathcal{F}\simeq1$, we have $D\simeq J_{\uparrow\uparrow}$, hence stabilizing the SS state. On the other hand for $k_0=0$, there is a cancellation of phases in the $\bm{k}$-sum in Eq.~\ref{eq:D}, suppressing $D$: the orbital content of the valleys winds in a manner that suppresses the $2\bm{q}_c$ CDW even when $\langle b^\dagger_+ b^{\phantom{\dagger}}_- \rangle\neq0$. $J$ generically remains non-zero, so if the Hartree cost for charge order (parametrized by $D$) is lowered sufficiently, the SDW can beat out the SS. 
Revisiting the HF numerics, we indeed find that the effective positions of the gapped Dirac points shift away from the minima towards $\bm{\Gamma}$ as $\epsilon^{-1}$ is increased, consistent with this scenario. This clarifies the relevance of Dirac and spin-orbit physics to the  excitonic order in WTe$_2$. 

\section{Collective modes} 
The distinct broken symmetries in the two excitonic phases lead to distinctive collective mode spectra~\cite{halperin1968long,nasu2016}. The three candidate  continuous global symmetries of the system (besides charge conservation, assumed throughout) are:  (i) the `excitonic' $U(1)_{eh}$ symmetry of treating electrons and holes (alternatively, conduction and valence electrons) as separately conserved~\cite{remez2020,murakami2020,golez2020}, generated by $a_\sigma \to a_\sigma e^{i\theta_{eh}}, b_{\sigma\lambda} \to b_{\sigma\lambda} e^{-i\theta_{eh}}$;   (ii) the $U(1)_v$ symmetry corresponding to independent conservation of electrons in the two valleys ($b_{
\sigma\lambda} \to b_{\sigma\lambda} e^{i\lambda\theta_v}$); and (iii) the $U(1)_s$ spin rotation symmetry about the $y$-axis, manifest in $H_0+H_{\text{int}}$ ($a_\sigma\to a_\sigma e^{i\sigma\theta_s}, b_{\sigma\lambda} \to b_{\sigma\lambda} e^{i\sigma\theta_s}$). The first two of these are explicitly broken at a microscopic level, since interactions mix bands, but are present in the DTA.  However as we show in Appendix~\ref{app:bDTA}, at bDTA level, interaction terms $\sim a^\dagger a^\dagger b_{+} b_{-}$  (in $\tilde{\mathcal{E}}[w]$) destroy the $U(1)_{eh}$ symmetry,  while $U(1)_v$ is preserved. Accordingly in the DTA/bDTA regime (throughout the SDW and in the SS for $\epsilon^{-1}\sim \epsilon^{-1}_{c1}$) we expect that interactions only weakly gap any $U(1)_v$ pseudo-Goldstone  modes but strongly gap the `excitonic' $U(1)_{eh}$ mode.  At this level, we take the $U(1)_s$ symmetry of $H_0+H_{\text{int}}$ to be exact, though as it is not symmetry-required it may be weakly broken 
 beyond the  $\bm{k}\cdot\bm{p}$ limit.
  Finally, since exciton condensation occurs at a  finite wavevector 
the breaking of  $U(1)_{s}$ and/or $U(1)_v$  is intertwined with translational symmetry breaking at $\bm{q}_c$. Hence we expect  a minimum (or gapless point) in the collective mode dispersion at  $\bm{q} \simeq \bm{q}_c$ in the original WTe$_2$ BZ.

The SDW breaks $U(1)_s$, with the corresponding freedom parameterized by $\phi$ in~\eqref{eq:SDW-w}, leading to a standard magnon mode (gapless in the interacting $\bm{k}\cdot\bm{p}$ model). In addition, since the conduction band minima are not at high-symmetry points and hence generically incommensurate with the undistorted lattice, we expect a phason mode (captured by $\alpha$) generated by $U(1)_v$ rotations, which will be weakly gapped as the latter is only an approximate symmetry. In contrast the SS only has a  phason mode, since spin and valley are locked. We anticipate this will again be weakly gapped,  with the gap increasing for higher interactions. Owing to the broken inversion symmetry in the SS, it also hosts domain walls separating regions with opposite handedness of spin rotation.   An out-of-plane magnetic field explicitly breaks $U(1)_s$; on symmetry grounds this allows SS to induce a small CDW  amplitude, which might be one route to its detection.
Since spectroscopy of collective modes can be a diagnostic of exciton condensation~\cite{kogar2017,murakami2020,remez2020,golez2020},  investigation of the collective excitation spectrum in WTe$_2$ is likely to be a fruitful avenue of study.

\section{Discussion} 
We have proposed that two distinct gapped excitonic phases can be generated by interactions in monolayer WTe$_2$, with one (SS) likely relevant to recent experiments~\cite{WTe2exciton}. Using an effective model, we have linked energetic competition between SS and a proximate SDW phase to the orbital structure of the renormalized bands near the Fermi energy. Since these depend on both the interaction strength and the initial semi-metallic Fermi surfaces --- which can be tuned  by adjusting the  interaction screening length and the electrostatic displacement field respectively --- it is possible that the SDW phase can be stabilized experimentally. The SDW and SS may be distinguished by their broken symmetries (especially the presence or absence of a $2\bm{q}_c$ charge modulation, and their distinct $\bm{q}_c$-spin orders) and the resulting differences in their collective excitations.

The unusual oscillations in high field  magnetoresistance~\cite{WTe2Landau}  occur on a {scale} ($\sim 100~\text{M}\Omega$) typical of insulators, yet their temperature dependence is not activated. The latter fact appears to rule out  explanations centred on  modulation of the excitonic gap~\cite{zhang2016,pal2016,lee2020}. Other proposed mechanisms for quantum oscillations in insulators~\cite{knolle2015,knollecooper2017,grubinskas2018} would manifest only in thermodynamic quantities but not magnetoresistance. A more exotic explanation invokes a fractionalized Fermi surface of neutral `composite exciton' quasiparticles, whose quantum oscillations can give a
 weak metallic contribution to charge transport superimposed on an activated background~\cite{chowdhury2018,sodemann2018}.  Given  the typical fragility of fractionalized phases, it seems unlikely to be energetically competitive at zero field with the large-gap broken-symmetry states found here. We therefore conjecture that if such a fractionalized phase exists, some yet-unknown mechanism must stabilize it at high fields. 
  Potential alternative explanations  invoking the field-induced CDW order in SS may also be interesting to pursue. Investigating the high-field phase structure is a subtle and urgent question for future work.

\begin{acknowledgements} We thank B.A.~Bernevig, Nick~Bultinck, L.~Fu, N.P.~Ong, Z.~Song, and Sanfeng~Wu for helpful discussions and correspondence. We acknowledge support from the European Research Council under the European Union Horizon 2020 Research and Innovation Programme, Grant Agreement No. 804213-TMCS (YHK, SAP). Additional support was provided by the Gordon and Betty Moore Foundation through Grant GBMF8685 towards the Princeton theory program. YHK acknowledges the hospitality of Princeton University.
\end{acknowledgements}

{
\appendix
\onecolumngrid

\section{$k\cdot p$ Model and Hartree-Fock (HF)}\label{app:HF}
The setup here closely follows the supplement of Ref.~\cite{WTe2exciton}. In the basis $\{\ket{d\uparrow},\ket{d\downarrow},\ket{p\uparrow},\ket{p\downarrow}\}$, the $k\cdot p $ Hamiltonian of monolayer WTe2 is
\begin{gather}
	H_0=\left(a\bm{k}^2+b\bm{k}^4+\frac{\delta}{2}\right)
	\begin{pmatrix}
		1 & 0 & 0 & 0 \\
		0 & 1 & 0 & 0 \\
		0 & 0 & 0 & 0 \\
		0 & 0 & 0 & 0
	\end{pmatrix}
	+\left(-\frac{\bm{k}^2}{2m}-\frac{\delta}{2}\right)
	\begin{pmatrix}
		0 & 0 & 0 & 0 \\
		0 & 0 & 0 & 0 \\
		0 & 0 & 1 & 0 \\
		0 & 0 & 0 & 1
	\end{pmatrix}
	+v_xk_x\tau_xs_y+v_yk_y\tau_ys_0\\
	a=-3,\quad b=18,\quad m=0.03,\quad \delta=-0.9,\quad v_x=0.5,\quad v_y=3
\end{gather}
where energies are measured in eV and lengths in \r{A}. The symmetries are inversion $\hat{P}=\tau_z$ and time-reversal $\hat{T}=is_y\hat{K}$, leading to a two-fold degeneracy of the bands under $\hat{P}\hat{T}$. With the above parameters, the bandstructure at charge neutrality consists of a hole pocket at the zone centre, and two electron pockets with minima at $\bm{q_c}=\pm 0.3144\hat{x}$. The undistorted lattice has reciprocal lattice vector lengths $G_x=1.81$ and $G_y=1.01$. The Fermi energy is $E_F\simeq -0.493$. Without the SOC term, the bandstructure contains two \emph{overtilted} Dirac cones at $\bm{q}_D=\pm 0.2469\hat{x}$. The $U(1)_s$ symmetric SOC term gaps the Dirac point, leading to an indirect negative band gap.

The interaction Hamiltonian is taken to be density-density in spin and orbital space
\begin{gather}
	\label{eq:HintSM} H_{\text{int}}=\frac{1}{2N\Omega}\sum_{\bm{k},\bm{p},\bm{q}}\sum_{\alpha,\beta}U(\bm{q})c^\dagger_{\bm{k}+\bm{q},\alpha}c^{\dagger}_{\bm{p}-\bm{q},\beta}c^{\phantom{\dagger}}_{\bm{p},\beta}c^{\phantom{\dagger}}_{\bm{k},\alpha}\\
	U(q)=\frac{e^2}{2\epsilon\epsilon_0 q}\tanh\frac{q\xi}{2}
\end{gather}
where $N$ is the total number of unit cells in the system, $\Omega$ is the real-space unit cell area, and $\alpha,\beta$ are combined orbital/spin indices. The interaction potential is of dual-gate screened form, with $\xi$ the gate distance and $\epsilon$ the relative permittivity of the encapsulating hBN. Experimentally relevant parameters are $\epsilon\simeq3.5$ and $\xi\simeq250$. Since we are working with a $k\cdot p$ model, our calculations require a momentum cutoff, which is taken to be $|k_x|<\frac{3q_c}{2},|k_y|<\frac{G_y}{4}$. The prefactor in Eq~\ref{eq:HintSM} is set by the density of momentum points, which would be $A_{BZ}/N$. However in our calculations our momentum cutoff has area $A_{kp}$ with $N_{kp}$ points, so we should replace $N\Omega\rightarrow N_{kp}\Omega\frac{A_{BZ}}{A_{kp}}$.

Anticipating excitonic pairing at $\pm \bm{q_c}$, we perform self-consistent HF calculations allowing for coherence by multiples of $\bm{q_c}$, i.e. $\langle c^\dagger_{\bm{k}\alpha}c^{\phantom{\dagger}}_{\bm{k}+n\bm{q_c}\beta}\rangle$ can take non-zero values for integer $n$. Representative (folded) band structures are shown in Figure~\ref{fig:U0bandstructures}. We allow the HF solution to break various symmetries such as time-reversal and inversion. To ensure that we find the lowest energy mean-field solution, we minimize over multiple random seeds for the initial density matrix. The presence of excitonic condensation is diagnosed by the integrated order parameter $\Delta_\text{exc}\equiv \frac{1}{N_{kp}}\sqrt{\sum_{\alpha,\beta} |\langle c^\dagger_{\bm{k}\alpha}c^{\phantom{\dagger}}_{\bm{k}+\bm{q_c}\beta} \rangle|^2}$. The spin-valley nature of the ordering is diagnosed by  computing (finite-momentum) charge/spin densities
\begin{gather}
	\rho^c_{\bm{Q}}=\frac{1}{\tilde{A}}\sum_{\sigma\bm{k}a}\langle c^\dagger_{\bm{k}-\bm{Q}\sigma a}c^{\phantom{\dagger}}_{\bm{k}\sigma a}\rangle\\
	\bm{\rho}^s_{\bm{Q}}=\frac{1}{\tilde{A}}\sum_{\sigma\sigma'\bm{k}a}\bm{\sigma}_{\sigma\sigma'}\langle c^\dagger_{\bm{k}-\bm{Q}\sigma a}c^{\phantom{\dagger}}_{\bm{k}\sigma' a}\rangle
\end{gather}
where $a$ runs over the orbital degree of freedom, and $\tilde{A}=N_{kp}\Omega\frac{A_{BZ}}{A_{kp}}$. Since there are no ferromagnetic states, the relevant densities for characterising the excitonic order are $\rho_{q_c}^c,\rho_{2q_c}^c,\bm{\rho}_{q_c}^s,\bm{\rho}_{2q_c}^s$. In real space, the number densities for spin and charge are
\begin{gather}
	\rho^c(\bm{r})=\sum_{\bm{Q}}e^{i\bm{q}\cdot\bm{r}}\rho^c_{\bm{Q}}\\
	\bm{\rho}^s(\bm{r})=\sum_{\bm{Q}}e^{i\bm{q}\cdot\bm{r}}\bm{\rho}^s_{\bm{Q}}.
\end{gather}
Self-consistent HF reveals that there are two energetically competitive types of excitonic insulator, the spin spiral (SS) and spin density wave (SDW) which both occur at $q_c$. Consider the magnitude of the local spin density in the presence of spin order which occurs at a single wavevector $q_c$
\begin{equation}
	\bm{\rho}^s(\bm{r})\cdot \bm{\rho}^s(\bm{r})=2\bm{\rho}^s_{q_c}\cdot\bm{\rho}^{s*}_{q_c}+(\bm{\rho}^s_{q_c}\cdot\bm{\rho}^s_{q_c}e^{iq_c x}+c.c.).
\end{equation}
Therefore the following quantities are useful to diagnose the presence of SS and SDW order
\begin{gather}
	\rho^\text{SDW}=\sqrt{2|\bm{\rho}^s_{q_c}\cdot\bm{\rho}^s_{q_c}|}\\
	\rho^\text{SS}=\sqrt{2\bm{\rho}^s_{q_c}\cdot\bm{\rho}^{s*}_{q_c}}-\sqrt{2|\bm{\rho}^s_{q_c}\cdot\bm{\rho}^s_{q_c}|}.
\end{gather}
A pure SS state has a local spin polarization that rotates with constant magnitude (comparing points with the same intra unit cell coordinates), while a pure SDW state has an oscillating local spin magnitude whose oscillation wavelength is longer than the monolayer unit cell.

\begin{figure}[!t]
	\includegraphics[width=1\linewidth,clip=true]{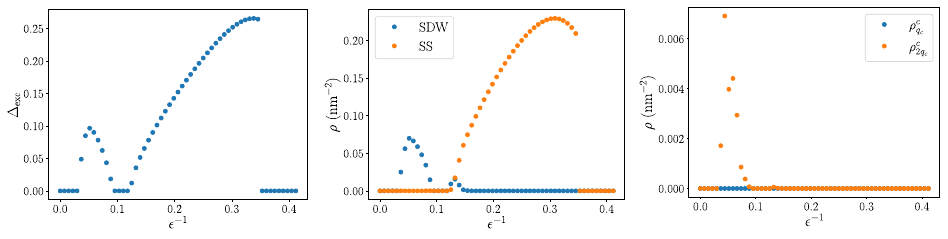}
	\caption{Self-consistent HF results for the excitonic, spin and charge order parameters as a function of interaction strength $\epsilon^{-1}$. Note that $\rho^{\text{SDW}}$ and $\rho^{\text{SS}}$ in the middle panel measure the average local spin population imbalance (number per square nanometer) due to the corresponding order.}
	\label{fig:orderparams}
\end{figure}
Self-consistent Hartree-Fock results for the various order parameters are shown in Figure~\ref{fig:orderparams}. Note that SDW order is accompanied by charge order at $2q_c$.

HF calculations were also performed without allowing for excitonic coherence, in order to obtain the `parent' self-consistent states appropriate for an analytic weak-coupling treatment of excitonic pairing. Representative results are shown in Figure~\ref{fig:parentbandstructures}. 

\begin{figure}[!t]
	\includegraphics[width=1\linewidth,clip=true]{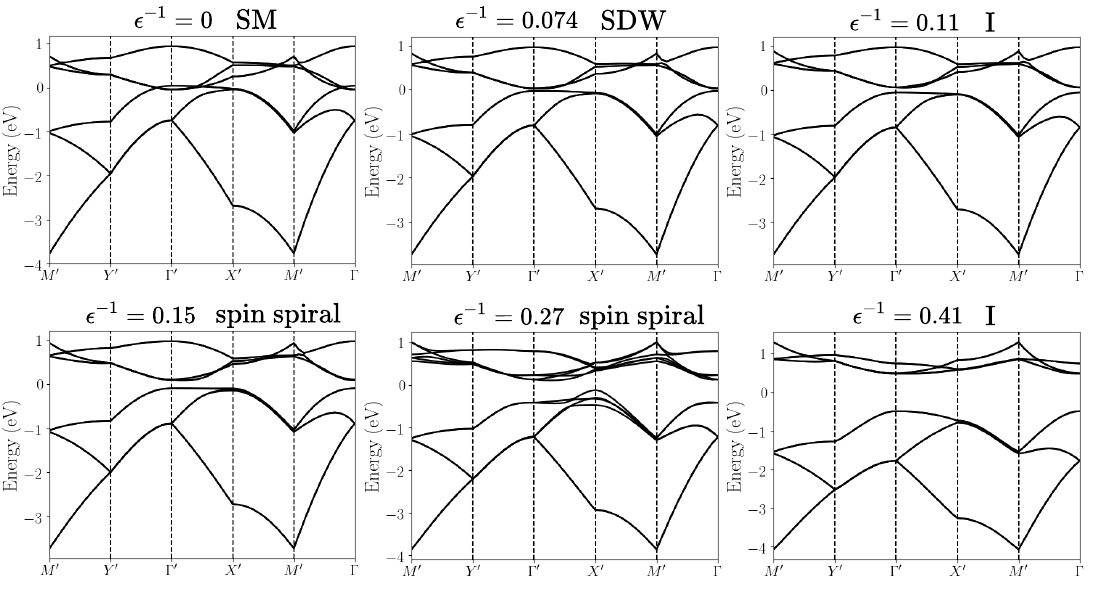}
	\caption{Folded HF band structures for different interaction strengths---in order of increasing $\epsilon^{-1}$, the phases are semimetal, SDW, non-excitonic quantum spin Hall insulator (diagnosed by using the Fu-Kane formula~\cite{FuKane}), spin spiral, spin spiral (at experimentally relevant $\epsilon^{-1}$), trivial insulator. Labeled momentum points are $M'=(-q_c/2,G_y/4),Y'=(0,G_y/4),X=(-q_c/2,0)$. Given the momentum cutoff of the $k\cdot p$ theory, there are 12 bands per momentum in the folded BZ. All phases except the spin spiral have doubly-degenerate bands due to $\hat{P}\hat{T}$ symmetry. Calculations were done on a $75\times25$ momentum grid.}
	\label{fig:U0bandstructures}
\end{figure}

\begin{figure}[!t]
	\includegraphics[width=1\linewidth,clip=true]{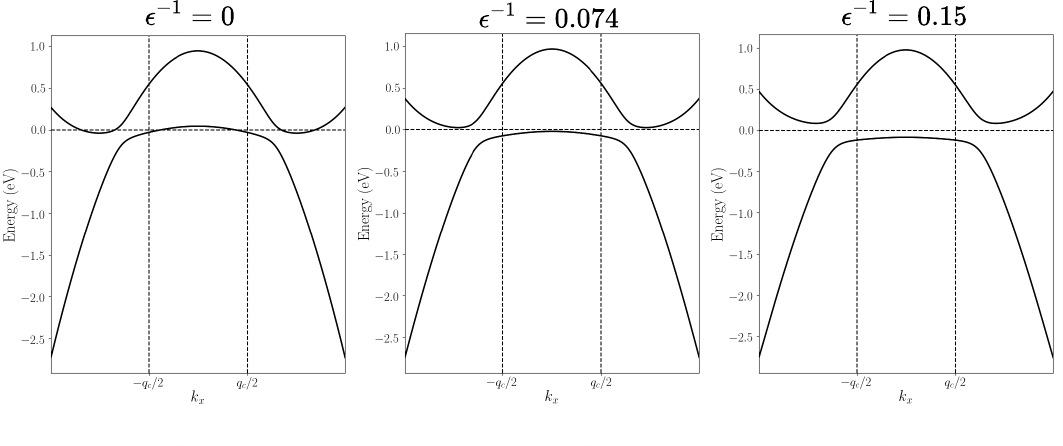}
	\caption{HF band structures along the $k_x$ axis, when the HF is restricted to forbid translation symmetry breaking. Calculations were done on a $75\times25$ momentum grid.}
	\label{fig:parentbandstructures}
\end{figure}

\section{Trial Excitonic Insulator States}
For generality, consider the situation with one valence pocket at $\bm{\Gamma}$, and $N_\lambda$ equivalent conduction valleys at $\bm{Q}(\lambda)$. The insulating excitonic trial states considered in the main text are of the following form
\begin{equation}\label{eq:varansatzSM}
	\ket{\Phi} = \prod_{\bm{k},\sigma} \alpha_{\bm{k}\sigma}^\dagger\ket{0},\,\alpha_{\bm{k}\sigma} =  u_{\bm{k}}a_{\bm{k}\sigma} + v_{\bm{k}}\sum_{s\lambda} w^{\sigma}_{s\lambda} b_{\bm{k}s\lambda},
\end{equation}
where $u_k,v_k$ are real and even, $u_k^2+v_k^2=1$, $\sum_{s\lambda}w^{\sigma}_{s\lambda}\bar{w}^{\sigma'}_{s\lambda}=\delta_{\sigma\sigma'}$, and overbar denotes complex conjugation. $u_k,v_k$ parameterizes the (small)-momentum structure of exciton coherence, while $w_{s\lambda}^\sigma$ parameterizes the spin-valley structure of pairing. Note that, for a fixed choice of gauge for the Bloch operators, the choice of orthonormal complex vectors $w^\uparrow,w^\downarrow$ in $\mathbb{C}^{2N_\lambda}$ uniquely specifies the trial state without any redundancy---there is no gauge redundancy corresponding to unitary rotation within occupied orbitals. 

Now specialize to the case of two valleys $\lambda=\pm$, so that $\bm{Q}(\lambda)=\lambda\bm{q_c}$. The trial states considered by Ref.~\cite{halperin1968long} are a strict subset of Eqn.~\ref{eq:varansatzSM}, and can be parameterized by
\begin{gather}
	w^\sigma_{s\lambda}=lM^{(+)}_{\sigma s}\delta_{\lambda +}+mM^{(-)}_{\sigma s}\delta_{\lambda -}
\end{gather}
where $l^2+m^2=1$, and the $M$ matrices are unitary. This can describe the SDW, but can only describe the spin spiral if $l$ and $m$ are allowed to be $\sigma$-dependent.

It can be shown that global spin-rotation $\hat{U}^s$ and valley-rotation $\hat{U}^v$ act as
\begin{align}\label{eqn:spinrotation}
	\hat{U}^s:&\quad w^\sigma_{s\lambda}\rightarrow \sum_{\sigma's'}U^\dagger_{\sigma\sigma'}w^{\sigma'}_{s' \lambda}U^{\phantom{\dagger}}_{s's}\\
	\hat{U}^v:&\quad w^\sigma_{s\lambda}\rightarrow \sum_{\lambda'}w^\sigma_{s\lambda' }U_{\lambda'\lambda}\label{eqn:valleyrotation}
\end{align}
where $U=\exp\left(\frac{i\theta}{2}\hat{n}\cdot\bm{\sigma}\right)$ is a $SU(2)$ unitary. $U(1)_{eh}$ rotations corresponding to separate conservation of conduction and valence populations act as $w^\sigma_{s\lambda}\rightarrow w^\sigma_{s\lambda}e^{i\theta}$.

\section{Dominant Term Approximation (DTA) Equations}\label{SecDTA}
The DTA equations~\cite{cloizeaux1965,halperin1968long} determine the internal momentum structure of excitonic coherence (i.e. the coefficients $u_k,v_k$). The starting point is an effective model that describes the band extrema of a self-consistent non-excitonic band structure, which can be semimetallic or insulating. For simplicity consider the `two-pocket' case (one conduction minimum $b^\dagger_{k\sigma}$ and valence maximum $a^\dagger_{ k\sigma}$)---the multi-valley case can be treated analogously. In the DTA, only intra-pocket interactions are included, and the gauge is chosen smooth so that $F^{nn;\sigma}_{k,k+q}\simeq1$ for small momentum transfer $q$, leading to the Hamiltonian
\begin{equation}
	\hat{H}_\text{DTA}=\sum_{kn\sigma}\epsilon^n_{k}d^\dagger_{nk\sigma}d_{nk\sigma}+\frac{1}{2}\sum_{kk'qnn'\sigma\sigma'}U(q)d^\dagger_{n,k+q,\sigma}d^\dagger_{n',k'-q,\sigma'}d^{\phantom{\dagger}}_{n',k',\sigma'}d^{\phantom{\dagger}}_{n,k,\sigma}.
\end{equation}
Therefore there is an emergent $U(1)$ symmetry corresponding to separate conservation of conduction and valence band electrons. There is also now global $SU(2)_s$ spin rotation symmetry, as well as $SU(2)_v$ valley rotation symmetry.

We consider an insulating excitonic ansatz $\ket{\Phi(w)}$  described by the operator for the filled bands
\begin{equation}
	\alpha_{k\sigma}=u_ka_{k\sigma}+v_k\sum_{s}w^\sigma_{s}b_{ks},\quad \sum_{s}w^{\sigma}_{s}w^{\sigma'^*}_{s }=\delta_{\sigma\sigma'}.
\end{equation}
where $u_k,v_k$ are real and even, and $u_k^2+v_k^2=1$. We now recall that the parameters of the model are extracted from a self-consistent band structure. Therefore when counting the interactions of any distorted state, we need to measure the density relative to the reference self-consistent state $\Phi_0$:
\begin{gather}
	E_\text{DTA}[\Phi(w)]=\text{const}+2\sum_{k}v_k^2(\epsilon^b_{k}-\epsilon^a_{k})-\frac{1}{2}\sum_{kk'nn'\sigma\sigma'}U(k-k')\langle d^\dagger_{nk'\sigma}d^{\phantom{\dagger}}_{n'k'\sigma'}\rangle'\langle d^\dagger_{n'k\sigma'}d^{\phantom{\dagger}}_{nk\sigma}\rangle'\\
	\langle a^\dagger_{k\sigma}a^{\phantom{\dagger}}_{k\sigma'}\rangle'=(u_k^2-N^0_{ak})\delta_{\sigma\sigma'}\\
	\langle b^\dagger_{k\sigma}a^{\phantom{\dagger}}_{k\sigma'}\rangle'=g_k w^{\sigma'}_{\sigma}\\
	\langle b^\dagger_{k\sigma}b^{\phantom{\dagger}}_{k\sigma'}\rangle'=(v_k^2-N^0_{bk})\delta_{\sigma\sigma'}.
\end{gather}
where $g_k=u_kv_k$, and $N^0_{nk}$ is the filling of $\Phi_0$ (for the insulating parent state in the main text, we have $N^0_{ak}=1$). The direct contributions with $q=0$ are canceled by the neutralizing background. Evaluating the interaction term, we obtain the DTA energy
\begin{equation}
	E_\text{DTA}=\text{const}+ 2\sum_{k}v_k^2(\epsilon^b_{k}-\epsilon^a_{k})-\sum_{kk'}U(k-k')\left[(u_k^2-N^0_{ak})(u_{k'}^2-N^0_{ak'})+(v_k^2-N^0_{bk})(v_{k'}^2-N^0_{bk'})+2g_kg_{k'}\right]
\end{equation}
which is independent of $w$, leading to a huge degeneracy at DTA level. We minimize this energy with respect to $v_k$
\begin{gather}
	0=\partial_{v_p}E_{\text{DTA}}=4(\epsilon^b_{p}-\epsilon^a_{p})v_p-4\sum_{k}U(k-p)\left[(2v_k^2-1+N^0_{ak}-N^0_{bk})v_p+g_k\frac{1-2v_p^2}{\sqrt{1-v_p^2}}\right]\\
	\rightarrow \left[\epsilon_{pb}-\sum_k U(k-p)\left(v_k^2-N^0_{bk}\right)-\epsilon_{pa}+\sum_k U(k-p)\left(u_k^2-N^0_{ak}\right)\right]v_p=\frac{1-2v_p^2}{\sqrt{1-v_p^2}}\sum_k U(k-p)g_k\\
	\rightarrow 2\xi_pv_p=\frac{1-2v_p^2}{\sqrt{1-v_p^2}}\Delta_p
\end{gather}
where we have defined
\begin{gather}
	\bar{\epsilon}^a_{k}=\epsilon_{ka}-\sum_{k'}U(k-k')(u_{k'}^2-N^0_{ak'})\\
	\bar{\epsilon}^b_{k}=\epsilon_{kb}-\sum_{k'}U(k-k')(v_{k'}^2-N^0_{bk'})\\
	\xi_k=\frac{1}{2}(\bar{\epsilon}^b_{k}-\bar{\epsilon}^a_{k})\\
	\Delta_k=\sum_{k'}U(k-k')g_{k'}.
\end{gather}
The minimization condition can be recast as the coupled integral equations
\begin{equation}
	v_k=\sqrt{\frac{1}{2}\left(1-\frac{\xi_k}{\sqrt{\xi_k^2+\Delta_k^2}}\right)}
\end{equation}
which are solved by iteration.

The energy bands of the excitonic state are given by 
\begin{gather}
	E_{k\alpha}=\frac{\bar{\epsilon}^a_{k}+\bar{\epsilon}^b_{k}}{2}-\sqrt{\xi_k^2+\Delta_k^2}\\
	E_{k\beta}=\frac{\bar{\epsilon}^a_{k}+\bar{\epsilon}^b_{k}}{2}+\sqrt{\xi_k^2+\Delta_k^2}.
\end{gather}
If we have two valleys, we will have an additional energy band $E_{k\gamma}=\epsilon^b_k$ which remains unaltered. In this case it is possible that the excitonic state remains semimetallic if the parent state is semimetallic.

\section{Beyond Dominant Term Approximation (bDTA) Splitting Terms}\label{app:bDTA}
While the DTA equations determine $u_k,v_k$, the choice of $w$ can only be resolved by considering the neglected inter-pocket interactions~\cite{halperin1968long}. Assuming a good DTA/bDTA separation of scales, we can use first-order perturbation theory to evaluate the neglected terms of $\braket{\Phi(w)|\hat{H}|\Phi(w)}$. In the two-valley case, we obtain
\begin{eqnarray}\label{eq:bDTASM}
	\delta E[w]&=&\sum_{\sigma\sigma'}\left(B_{\sigma\sigma'}(w^\sigma_{\sigma+}+\bar{w}^{\bar{\sigma}}_{\bar{\sigma}-})(\bar{w}^{\sigma'}_{\sigma'+}+w^{\bar{\sigma'}}_{\bar{\sigma'}-})-2\text{Re}\,C_{\sigma\sigma'}w^{\sigma'}_{\sigma+}w^{\sigma}_{\sigma'-}\right)\\
	&+&\sum_{\sigma\sigma'ss'}\left(Dw^\sigma_{s+}\bar{w}^{\sigma}_{s-}\bar{w}^{\sigma'}_{s'+}w^{\sigma'}_{s'-}-J_{ss'}w^\sigma_{s+}\bar{w}^{\sigma}_{s'+}w^{\sigma'}_{s'-}\bar{w}^{\sigma'}_{s-}\right)
\end{eqnarray}
\begin{eqnarray}\label{eq:bDTAcoeffs}
	B_{\sigma\sigma'}&=&U(q_c)\sum_{k}g_kF^{ab;\sigma^*}_{k,k+q_c}\sum_{k'}g_{k'}F^{ab;\sigma'}_{k',k'+q_c}\\
	C_{\sigma\sigma'}&=&\sum_{kk'}g_kg_{k'}U(k-k'+q_c)F^{ab;\sigma^*}_{k',k+q_c}F^{ba;\sigma'}_{k'-q_c,k}\\
	D&=&U(2q_c)|\sum_{k}v_k^2F^{bb;\sigma}_{k-q_c,k+q_c}|^2\\
	J_{ss'}&=&\frac{1}{2}\sum_{kk'}v_k^2v_{k'}^2U(k-k'+2q_c)[F^{bb;s^*}_{k'-q_c,k+q_c}F^{bb;s'}_{k'-q_c,k+q_c}+\text{c.c.}].
\end{eqnarray}
where $B$ is Hermitian, $J,C$ are symmetric, and the spin-quantization axis is chosen along the preserved direction (SOC is $U(1)_s$ preserving). The $F$ refer to the form factors of the effective model, and the momentum labels are absolute momenta measured from the zone center. For example, $F^{bb;\sigma}_{k-q_c,k'+q_c}$ is an intervalley form factor because the $k,k'$ always represent `small' momenta. The $B$-term and $D$-term are Hartree terms that penalize charge density wave modulations at wavevector $q_c$ and $2q_c$ respectively. The $C$-term and $J$-term are exchange terms at momentum transfer $q\sim q_c$ and $2q_c$ respectively. In the limit of vanishing SOC, and specializing to a restricted class of states (that includes the SDW but not the spin spiral), we recover the bDTA expression of Ref.~\cite{halperin1968long}.

Using the transformations in Eqns~\ref{eqn:spinrotation},\ref{eqn:valleyrotation}, it can be shown that $\delta E[w]$ is invariant under $U(1)_s$ and $U(1)_v$ symmetries. $U(1)_s$ is present because the starting model was already assumed to have this symmetry. $U(1)_v$ can be seen by investigating the possible inter-pocket interaction terms which conserve momentum. This symmetry ceases to be sensible once excitonic coherence remains strong out to momenta $k\sim q_c/2$ in the folded BZ, since then the division of the relevant low-energy Bloch states into small `pockets' fails, and the weak-coupling perspective is no longer useful. There is no $U(1)_{eh}$ symmetry corresponding to separate conservation of valence/conduction electrons, because bDTA contains interaction terms $\sim a^\dagger a^\dagger b^{\phantom{\dagger}}_+b^{\phantom{\dagger}}_-$. These $U(1)_{eh}$-violating terms are reflected in the $B$- and $C$-terms of the bDTA energy functional.

\section{Spiral/SDW Competition for Two Excitons}\label{app:twoexcitons}
In this section we argue that the spin spiral vs SDW competition outlined in the main text is invisible to a single exciton, and is a selection mechanism at the many-exciton level. Let $\ket{\Phi_0}=\prod_{k\sigma}a^\dagger_{k\sigma}\ket{\text{vac}}$ be the parent insulating state of the effective model. An exciton creation operator (with net momentum $q=0$ in the folded BZ) can be parameterized as a linear combination of single particle-hole operators:
\begin{equation}
	B^\dagger_\sigma(w)=\sum_{ks\lambda}f_k w^*_{s\lambda}b^\dagger_{ks\lambda}a^{\phantom{\dagger}}_{k\sigma}
\end{equation} 
where $f_k$, satisfying $\sum_k f_k^2=1$, is real and even, and parameterizes the exciton momentum structure (predominantly determined by $q\sim 0$ interactions), while $w$ indicates the valley/spin structure of the electron. 

We focus on the $q\sim 2q_c$ components of the interaction Hamiltonian, since these were found to mediate the SDW/spiral competition. Consider a single exciton state
\begin{equation}
	\ket{\sigma w}=B_\sigma^\dagger(w)\ket{\Phi_0}.
\end{equation}
This vanishes under the action of $q\sim 2q_c$ interaction terms, since $b^\dagger b^\dagger bb$ will always annihilate the above state. Hence a single exciton is not sensitive to the competition described in the main text.

Now we consider the interaction energy of two-exciton states. For simplicity we assume the quasi-boson approximation and consider the following two-exciton states
\begin{equation}
	\ket{\sigma w;\sigma' w'}=B^\dagger_\sigma(w)B^\dagger_{\sigma'}(w')\ket{0}
\end{equation}
where we neglect the normalization. We will be interested in cases where the $w,w'$ describe excitons with spin/valley structures corresponding to spiral or SDW phases. Focusing on the $q\sim2q_c$ contributions again, we obtain after some algebra
\begin{align}
	\braket{\sigma w;\sigma w'|\hat{H}_{q\sim2q_c}|\sigma w;\sigma w'}&=\frac{2}{N}\bigg[U(2q_c)|\sum_k f_k^2\mathcal{F}^{\uparrow}_{k,k}|^2\sum_{ss'}(w_{s+}\bar{w}_{s-}\bar{w}'_{s'+}w'_{s'-}+\text{c.c.})\\
	&-\frac{1}{2}\sum_{kk'ss'}f_k^2f_{k'}^2U(k-k'+2q_c)(\mathcal{F}^{s^*}_{k'k}\mathcal{F}^{s'}_{k'k}w_{s+}\bar{w}_{s'+}w'_{s'-}\bar{w}'_{s-}+\text{c.c.})\bigg].
\end{align}
For $w,w'$ corresponding to the spin spiral (Eqn 10 in main text), the above contributions vanish as expected since there is no $2q_c$ coherence. For $w,w'$ corresponding to the SDW (Eqn 9 in main text), we recover the competition between the direct and exchange terms, which take analogous forms to the $D$- and $J$-terms in the bDTA. We note that these calculations (involving $q\sim0$ and $q\sim q_c$ terms as well) can be generalized to derive the interaction terms of an effective quasi-boson Hamiltonian.

\section{Elliptical Spin Spirals and Charge Order}\label{app:elliptical}
In this section we consider the more general class of elliptical spin spirals, which encompasses the limiting cases of SDW and (circular) spin spiral discussed in the main text. The spin/valley structure of these states can be parameterized using the language of Eq~\ref{eq:varansatzSM}
\begin{align}
	w^{\uparrow}_{\downarrow+}&= e^{-i(\alpha+\phi)}\sin(\chi-\frac{\pi}{4})\\
	w^{\uparrow}_{\downarrow-}&= e^{i(\alpha-\phi)}\cos(\chi-\frac{\pi}{4})\\
	w^{\downarrow}_{\uparrow+}&= e^{i(-\alpha+\phi)}\cos(\chi-\frac{\pi}{4})\\
	w^{\downarrow}_{\uparrow-}&= e^{i(\alpha+\phi)}\sin(\chi-\frac{\pi}{4}).
\end{align}
The spin spiral is recovered for $\chi=\pm\pi/4$ (corresponding to the two senses of rotation), while $\chi=0$ corresponds to the SDW. The spin and charge densities for these states are 
\begin{gather}
	\bm{\rho}^s(\bm{r})\sim\begin{bmatrix}
		\sin\chi\cos\phi\cos\big[q_cx+\alpha\big]+\cos\chi\sin\phi\sin\big[q_cx+\alpha\big]\\
		0\\
		\sin\chi\sin\phi\cos\big[q_cx+\alpha\big]+\cos\chi\cos\phi\sin\big[q_cx+\alpha\big]
	\end{bmatrix}\\
	\rho^c(\bm{r})\sim\cos2\chi\cos\big[2(q_cx+\alpha)\big].
\end{gather}
Hence the principal axes of the elliptical spiral are controlled by $\phi$ and lie along $[\cos\phi,0,\sin\phi]$ and $[-\sin\phi,0,\cos\phi]$, while $\alpha$ controls the position along $x$. $\chi$ is related to the ellipticity of the spin order, and also controls the strength $\sim\cos^2 2\chi$ of the associated $2q_c$ charge density wave. 

To further understand the relation between the ellipticity of the spiral and the charge order, we can analyze their coupling within Landau theory~\cite{Zachar}. With the constraints given by TRS, $U(1)_s$ about $s^y$, and translation (there are no Umklapp processes since $\bm{q_c}$ is not at a high-symmetry point), the lowest order coupling between charge density $\rho^c$ and $x-z$ spin density $\bm{\rho}^{s,\perp}$ is
\begin{align}
	F\sim \int dx \rho^c(x)\big[\bm{\rho}^{s,\perp}(x)\big]^2
	\sim \sum_{p,p'}\rho^c_{-p-p'}\bm{\rho}^{s,\perp}_p\cdot\bm{\rho}^{s,\perp}_{p'}
\end{align}
where we have used the fact that there is no order along the $y$-direction. Since the spin order has non-trivial contributions for momenta $p=\pm q_c$, we focus on the case $p=p'$ which couples to the $2q_c$ charge order. With appropriate choice of coordinate and spin axes, the spin order parameter of the elliptical spiral can be chosen as $\bm{\rho}^{s,\perp}(x)\sim[\cos\chi\sin(q_c x),\sin\chi\cos(q_c x)]$, with Fourier components $\bm{\rho}^{s,\perp}_{\pm q_c}\sim[\pm i\cos\chi,\sin\chi]$, leading to $\bm{\rho}^{s,\perp}_{\pm q_c}\cdot\bm{\rho}^{s,\perp}_{\pm q_c}\sim \cos2\chi$. Hence the coupling between $2q_c$ charge order and $q_c$ spin order contains a multiplicative factor of $\cos2\chi$. This vanishes for the circular spiral, which can be intuited from the fact that $\big[\bm{\rho}^{s,\perp}(x)\big]^2$ is spatially uniform. This argument holds for higher order terms in Landau theory, since in-plane spin must enter as $\big[\bm{\rho}^{s,\perp}(x)\big]^2$ due to $U(1)_s$.

In the bDTA, the energy of the elliptical spiral is $E_\text{bDTA}=-2\text{Re}\,C_{\uparrow\downarrow}+\cos^22\chi\left(D-\frac{J}{2}\right)$. Hence the energetics mean that we have $\chi\rightarrow \pm\pi/4,0$ depending on whether $D-\frac{J}{2}$ is positive or negative.	

\textbf{ $U(1)_s$ symmetry breaking and CDW modulation in a circular spiral phase:} Note that when $U(1)_s$ symmetry in the $xz$ spin plane is broken (e.g., by a  magnetic field perpendicular to the monolayer), the reduction in symmetry admits additional terms such as $({\rho}^{s,x}{\rho}^{s,x} - {\rho}^{s,z}{\rho}^{s,z})\rho^c$, allowing even a circular spiral to generate a CDW modulation.

\clearpage
}

\twocolumngrid

\end{document}